\documentclass[letterpaper, 10 pt, conference]{ieeeconf}  

\IEEEoverridecommandlockouts                              
\usepackage{epsfig} 
\usepackage{mathptmx} 
\usepackage{times} 
\usepackage{hyperref}
\usepackage{amsmath,amssymb,amsfonts}
\usepackage{algorithmic}
\usepackage[euler]{textgreek}
\usepackage{multirow}
\usepackage{wrapfig}
\usepackage{subfigure} 
\usepackage[colorinlistoftodos]{todonotes}

\title{\LARGE \bf
\emph{``What's ur type?"}\\
Contextualized Classification of User Types in Marijuana-related  Communications using Compositional Multiview Embedding
}

\author{Ugur Kursuncu$^{1,3}$, Manas Gaur$^{1}$, Usha Lokala$^{1}$, Anurag Illendula$^{1}$, Krishnaprasad Thirunarayan$^{1}$, \\Raminta Daniulaityte$^{1,2}$, Amit Sheth$^{1}$ \& I. Budak Arpinar$^{3}$
\thanks{$^{1}$ Kno.e.sis Center, Wright State University, Dayton, OH, USA
	{\tt\footnotesize \{ugur,manas,usha,anurag,tkprasad,amit\}@knoesis.org}
    }%
\thanks{$^{2}$ CITAR, Wright State University, Dayton, OH, USA
{\tt\footnotesize raminta.daniulaityte@wright.edu}
    }%
\thanks{$^{3}$ Dept. of Computer Science, University of Georgia, Athens, GA, USA
	{\tt\footnotesize \{kursuncu,budak\}@uga.edu}
    }%
}

\begin{document}

\maketitle
\thispagestyle{empty}
\pagestyle{empty}

\begin{abstract}

With 93\% of pro-marijuana population in US favoring legalization of medical marijuana\footnote{\url{https://goo.gl/um2dDE}}, high expectations of a greater return for Marijuana stocks\footnote{\url{https://goo.gl/spSnVX}}, and public actively sharing information about medical, recreational and business aspects related to marijuana, it is no surprise that marijuana culture is thriving on Twitter. After the legalization of marijuana for recreational and medical purposes in 29 states\footnote{\url{https://goo.gl/CJVX5a}}, there has been a dramatic increase in the volume of drug-related communication on Twitter. Specifically, Twitter accounts have been established for promotional and informational purposes, some prominent among them being  \emph{American Ganja, Medical Marijuana Exchange}, and \emph{Cannabis Now}. Identification and characterization of different user types can allow us to conduct more fine-grained spatiotemporal analysis to identify dominant or emerging topics in the echo chambers of marijuana-related communities on Twitter. In this research, we mainly focus on classifying Twitter accounts created and run by ordinary users, retailers, and informed agencies.  Classifying user accounts by type can enable better capturing and highlighting of aspects such as trending topics, business profiling of marijuana companies, and state-specific marijuana policymaking. Furthermore, type-based analysis can provide more profound understanding and reliable assessment of the implications of marijuana-related communications. We developed a comprehensive approach to classifying users by their types on Twitter through contextualization of their marijuana-related conversations. We accomplished this using compositional multiview embedding synthesized from People, Content, and Network views achieving 8\% improvement over the empirical baseline.\\

\vspace{-1em}
\begin{keywords}
Semantic Social Computing, Compositional Multiview Embedding, Emoji Embedding, Network Embedding, Marijuana, User classification
\end{keywords}

\end{abstract}

\section{INTRODUCTION}
\label{sec:I}
\textit{``It's 4/20, and that means everyone is talking about marijuana\footnote{\url{https://goo.gl/JGSs3X}},"} highlights the state of marijuana-related communication on Twitter, especially around the time marijuana legalization polls were conducted in the USA. As more evidence is gathered through research studies on the safety and benefits of the medical and recreational uses of cannabis, there is a rise in public demand for broader legalization of marijuana and its variants. Accordingly, it is useful to study the engagement of users on social media to understand public opinion and its influence on policies better. 

Characterization of marijuana concentrate users on social media can enable researchers and analysts to describe the patterns of use, reasons of symptoms, and side effects as well as identify the predictor of risks with the help of spatiotemporal analysis. Specifically, classification of user types in marijuana communications on social media can aid in analyzing content-network dynamics at a user level, through an assessment of homophily in marijuana-related communities. Further, assessing the differences concerning marijuana conversations, the information flow, and interactions between user types, such as retail, informed agency and personal accounts, can help better situate their characteristics and understand the implications. For instance, in the case of predicting the outcome of a state legalization process\cite{kursuncu}, understanding the public opinions of the residents, assessing trending marijuana related topics in their conversations and monitoring their implications, are relevant and critical, as these opinions translate to votes. We associate personal user type (P) with an account handled by an individual user expressing their opinions, retail user type (R) with an account managed by a business entity to promote and market marijuana-related products, and informed agency user type (I) with an account handled by a group or organization to disseminate marijuana related information. Throughout the paper, we use informed agency \& media interchangeably to refer to the same user type.

In this study, we are proposing a user classification approach exploiting the multiview aspect of the Twitter data and features extracted from people, content, and network dimensions. The multiview stems from the inclusion of text, image (profile pictures), emoji and network interactions among accounts of different user types \cite{benton2016learning}.  Hence, for a reliable classification, we use \emph{Compositional Multiview Embedding}(CME) that combines different elements of the context such as text, image, emoji and network activities. 

This study addresses two key challenges: (i) The imbalanced dataset due to the relatively few users pertaining to Retail and Informed Agency user types, and (ii) Lack of proper use of different contextual dimensions, precisely, by incorporating Person-Content-Network views in compositional multiview embedding, for interpreting marijuana-related Twitter data.  

We create compositions of vector embeddings of these views of the Twitter data, called \emph{Compositional Multiview Embedding}(CME),  as it can represent the context in a more coherent manner \cite{mitchell2010composition}. In our approach, we create two CMEs: (i) one using tweet text, emoji and network interactions of users, and (ii) another using user description and emoji. In Section \ref{sec:CA}, we explain the correlation analysis performed on various feature combinations to assess their relationship. For instance, we found that descriptions and network interactions of users are highly correlated, suggesting that their combination can affect the performance of the classifier over the validation and test data. Therefore, we did not create the embedding using these two views. We evaluated the classifiers based on the individual F-scores of user type classes. We also generated word embedding vectors for profile pictures of users, which significantly improved the performance of classification of the informed agency user type. Details of our approach and results are discussed in Sections \ref{sec:M} and \ref{sec:R} respectively.

The remainder of the paper is organized as follows: In Section \ref{sec:RW}, we explain related works on the marijuana-related user classification. In Section \ref{sec:P}, we provide preliminaries about the concepts and technologies used. In Section \ref{sec:eda}, we provide an exploratory analysis that includes statistics about our dataset. Section \ref{sec:M} explains features and our experimental settings, and Section \ref{sec:R}  discusses the results of our analysis. Section \ref{sec:C}  concludes the paper with a summary and future research directions.

\section{RELATED WORK}
\label{sec:RW}
In this section, we describe prior studies that are broadly related to user classification, under three prominent sub-headings: (i) Embedding based Approaches to User Classification,  (ii) Diverse Features for User Classification, and (iii) User-level Approaches. 
\vspace{-0.6em}
\subsection{Embedding based Approaches to User Classification}
The profile of a user on Twitter consists of user description, tweets exchanged with their followers/friends,  profile picture. Researchers \cite{zhang2016user} utilized user tweets to learn an embedding model using Long Short-Term Memory (LSTM) and Recurrent Neural Network (RNN) to classify users based on their gender and age information achieving an accuracy of 91\% and 82\% respectively. In contrast, \cite{rizoslearning} employed interactional features to generate embeddings for a semi-supervised approach. Specifically, they utilize a small number of seed users with labels (e.g., news agency, person, genres) and interactions with ``mentions'' in their tweets. \cite{liao2017attributed} proposed an approach to learn the interactional features of the users by optimizing the structural and attribute level properties of their networks that characterizes homophily in their communication. In another study \cite{benton2016learning}, researchers utilized person-level multiview embedding to predict engagement, friend selection and demographic information of users. In contrast, our study gleans person, content and network-level features, creating a composition of multiview embeddings through \textit{vector addition} operation that characterizes users in the context of marijuana-related communications on social media.  
\vspace{-0.8em}
\subsection{Diverse Features for User Classification}
Prior work related to user classification on social media has involved different sets of features. Person-level features included profile \cite{pennacchiotti2011democrats}, user behavior, first and last names \cite{bergsma2013broadly}, demographics, Content level features including linguistic, domain-specific and generic LDA topics, and Network level features comprising follower-followee connections \cite{pennacchiotti2011democrats}. These features were utilized to glean political affinity, ethnicity, and favorability towards a particular profession, to generate machine-readable user-profiles for improving the user classification \cite{pennacchiotti2011democrats}, and to cluster users based on their conversations and predict demographics \cite{bergsma2013broadly}. Combination of these features with network interactions results in a better-contextualized representation of the dataset \cite{campbell2013content+}. They claimed that their model provides an in-depth analysis of users' communication from both content and network perspective, and improves user classification.  
\vspace{-0.7em}
\subsection{User-level Approaches}
For particular problems such as identification of user interests and event detection, user-level understanding of the content as well as the network dynamics is pivotal. In \cite{de2012unfolding}, they classified users into three classes, namely, organization, journalist (or media personnel), and an ordinary person, to identify variation in characteristics across multiple events. Engagement of users on a particular subject on social media is considered as an important signal in social media analytics, and has been used for user classification in \cite{tinati2012identifying}. The authors developed a working model to categorize a user as Idea Starter, Commentator, Curator, Amplifier, and Viewer. In the election domain, political homophily on social media forms a feature for user classification,  and \cite{colleoni2014echo} illustrates its significance for resolving reciprocated or non-reciprocated ties in the network of users. Homophily creates social echo chambers which polarizes the world of users. This fact  can be used to discriminate ordinary users  (or information seekers) from information providers (e.g., journalist). Moreover, topical analysis of the user-generated content can be informative about their intentions. In \cite{fang2015topic}, topic-centric Naive Bayes classifier was developed to identify the topics to categorize unknown users based on the closeness of their topics to those of the users in the training dataset.   
In recent years, there has been a surge in the use of e-cigarettes among smokers, and Twitter has emerged as a cost-effective platform for sharing and promoting information. In \cite{kim2017classification}, a user classification approach has been designed employing metadata and tweeting behavior to classify users as individuals, informed agencies, marketers, spammers, and vapor enthusiasts.

\section{PRELIMINARIES}
\label{sec:P}
Our approach uses several building blocks for an in-depth analysis of tweet content to extract relevant context in marijuana dataset.  Specifically, we discuss the people-content-network paradigm \cite{purohit2011understanding} and compositional word embeddings for expressiveness, EmojiNet for interpreting emoji, Clarifai for processing profile pictures, and SMOTE for oversampling. 
\vspace{-0.6em}
\subsection{People-Content-Network}
On social media, communities are being formed around various topics of interest through network interactions \cite{purohit2011understanding}. The information being shared in tweets by a user in the marijuana community displays an intent that depends on the user's type \cite{purohit2015intent}. For instance, \textit{personal users} share their experiences and opinions on marijuana, \textit{retail accounts} usually promote the use of marijuana and other related products that they sell, and \textit{media accounts} disseminate information on marijuana-related events and festivals,  legalization processes. Accordingly, as these user types show different characteristics, it is critical to bring to bear different perspectives, such as person, content, and network, for reliable analysis and insights. We describe a systematic organization and analysis of features in  Section \ref{sec:FE}. 
\vspace{-0.6em}
\subsection{EmojiNet}
\label{sec:enet}
Emoji are a pictorial representation of facial expressions, places, foods and other objects. These are often used by marijuana community on social media to express opinions and emotions about marijuana-related topics. Emoji contribute to the interpretation of the content created by the users and can contribute to the better recognition of the characteristics of user types. To achieve this goal, we make use of EmojiNet \cite{wijeratne2017emojinet} which gathers meanings of 2,389 emoji. Specifically, EmojiNet provides a set of words (e.g., smile), associated POS tags (e.g., verb), and their sense definitions. It maps 12,904 sense definitions to 2,389 emoji, to capture platform-specific interpretations. 

\vspace{-0.6em}
\subsection{Word Embedding Model}

A word embedding model created using word2vec can learn a rich low dimensional representation of words in a tweet corpus. Initially, the word embedding procedure was developed to generate distributional representations over corpora such as Wikinews, News articles, and Google News corpus that represent the current state-of-the-art. \cite{mikolov2013distributed} also shows that vector arithmetic over the word vectors can be used to generate analogies. For instance, word embedding of "Queen" can be obtained by summing the word embeddings of  "Man" and "Woman" and subtracting from it the word embedding of "King." 


In recent studies \cite{lilleberg2015support, wang2016semantic}, the researchers have shown that word embedding models perform well over short texts. In another study \cite{godin2015multimedia}, the authors have created a “named entity recognition shared task” for data from microblogging platforms using distributed word representations. These recent and prior successes in modeling words as computable vectors have encouraged us to utilize a pre-trained word2vec model trained over a generic Twitter corpus \cite{godin2015multimedia}  or train a new word-embedding model over our domain-specific Twitter corpus. Depending on the type of the corpus (characterized using sentence level statistics and word frequency counts), we can use one of two neural network architectures for learning word2vec embeddings: (i) Continuous Bag-Of-Words model \cite{mikolov2013distributed} (CBOW) (ii) Skip-gram model \cite{mikolov2013distributed}. In our study we have used skip-gram architecture. 

\vspace{-0.6em}
\subsection{Compositional Word Embedding}

In our study, we utilize compositional word embedding \cite{mitchell2010composition} to combine feature-level embedding vectors and to generate a comprehensive representation of a data point (e.g., user, tweet, user descriptions). Specifically, we employ weighted vector addition, a linear composition function detailed in \cite{mitchell2010composition}. Formally, we define Z, the weighted composition of word embeddings of U and V as follows: 
$\mathbf{Z} = W_0 \cdot U + W_1 \cdot V$,     
where $\mathbf{U,V} \in \mathbb{R}^{m\times300}$($\mathbf{m}$ represent number of users) are two embeddings  which are composed by weight-based (e.g., cosine similarity matrix) modulation using $\mathbf{W_0, W_1} \in  \mathbb{R}^{m\times m}$, respectively.  Note that in such a composition, the dimension of input and output representation is unaltered.
As detailed in Section \ref{sec:CA},  it is essential to consider the correlation between different view embeddings before composing them. For instance, in $\mathbf{Z}$ the weight matrices will be optimized through an optimization function; however, if the embeddings U and V are uncorrelated, it is computationally hard to generate the representation of $\mathbf{Z}$ as such optimization function over the two uncorrelated embeddings, will fail to converge.
Hence, we performed a linear composition, vector addition, to generate the representation of $\mathbf{Z}$. Since the classification is insensitive to the position of emoji and words in the content, we consider such composition as appropriate.
Formally, $\mathbf{Z}$= U+V is a vector addition of U and V. 



\section{EXPLORATORY ANALYSIS}
\label{sec:eda}
\begin{figure}
	\vspace{-0.5em}
    \centering
   \includegraphics[width=0.25\textwidth, angle=-90]{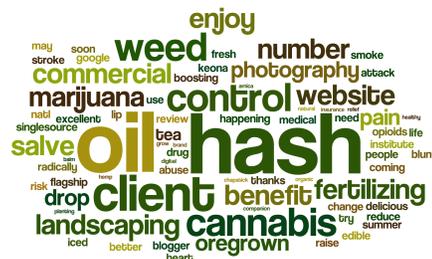}
    \vspace{-2em}
    \caption{\footnotesize Word Cloud of  tweets from Informed Agency.}
    \label{fig:iwc}
 \vspace{-2em}
\end{figure}

We have conducted an analysis of our dataset by extracting statistical, textual and topical information. 
Fig \ref{fig:iwc} captures the word cloud synthesized using the tweets from the Informed Agency user type that can be used to glean related topics. 
 
We have three classes of user types, namely, Personal Accounts (P), Informed Agency (I), and Retail Accounts (R). 
Our corpus contains tweets crawled in Summer 2017 that includes the months of June, July, and August, covering all states in the U.S.  During this time frame, the volume of communication-related to marijuana was high due to ongoing events (e.g. Cannabis Cup, The 420 Games)\footnote{\url{https://goo.gl/Xzsd5V}}. Data collection involved semantic filtering \cite{sheth2016semantic} utilizing the DAO\footnote{\url{https://goo.gl/9cXQcT}} ontology on the eDrugTrends\footnote{\url{http://wiki.knoesis.org/index.php/EDrugTrends}}/Twitris platform\footnote{\url{http://twitris.knoesis.org}}. The corpus comprised of a total of 4,106,566 tweets from 1,066,615 unique users. Out of nearly 4.1M tweets, 1,895,777 tweets were identified as unique through tweet id and the content.

\begin{table}[!htbp]
\vspace{-1em}
\scriptsize
\centering
\caption{\scriptsize Descriptive Information on the Training Set. \scriptsize}
\vspace{-1em}
\begin{tabular}{|p{3.2cm}|c|c|c|c|}
\hline
\textbf{Features ('\#' is ``number of")} & \textbf{P}& \textbf{R} & \textbf{I} & \textbf{Total} \\ \hline
\#Tweets (T) & 9836 & 1928 & 338 & 12102 \\ \hline
\#Profile Pictures (PP) & 4394 & 476 & 111 & 4981 \\ \hline
\#Users use Emoji (E) & 1085 & 37 & 17 & 1139 \\ \hline
\#Users with Descriptions (D) & 3884 & 461 & 108 & 4453 \\ \hline
\#Retweets & 955 & 24 & 964 & 1943 \\ \hline
\#Mentions & 94 & 6 & 307 & 407 \\ \hline
\end{tabular}
\label{tab:desc}
\vspace{-1.2em}
\end{table}

We randomly selected a set of 4982 users with 12,103 tweets from our pool of ~1M unique users to be considered as the training set. The domain experts from CITAR\footnote{\url{https://medicine.wright.edu/citar}} annotated the 4982 users in our training dataset as one of the following three types: Personal Accounts, Informed Agency, and Retail Accounts. After the annotation process, the distribution per user type was as follows: 4395 personal, 476 informed agency, and 111 retail accounts. Effectively, the distribution of user types in the training set is highly skewed. The reason for sparsity among retailers (i.e., retail business twitter accounts) is that marijuana is a schedule I\footnote{\url{https://goo.gl/UQhR4D}} drug according to the federal law, and thus its promotion of social media platforms is complicated due to its federal status as an illegal drug. 
Similarly, media accounts are significantly smaller compared to personal accounts, but still significantly higher than the retail accounts. Such data imbalance poses a serious risk in biasing the classifier towards the majority class. 

Upon our initial exploratory analysis of the corpus, we saw that the content in tweets and description of users are adequate to identify the characteristics of different user types. The average number of words in descriptions and tweets are 9.6 and 12.8, while the average number of emoji in descriptions and tweets are 0.46 and 0.26, respectively. 88\% of the users have their descriptions complete, and these user descriptions carry information containing emoji and text that can be utilized for classification. 

Further, interactions among users can play an essential role in disseminating the information and influence other connected users in the network. The median number of followers and friends for users are 367 and 376 respectively, and the average number of tweets per user is 3.85. Our corpus includes 2,837,734 interactions (mentions, retweets) between users, 83\% of which are retweets, and the rest are mentions. This statistics suggests that there is much communication among users that can contribute to the classification of user types.

\vspace{-0.5em}

\section{METHODOLOGY}
\label{sec:M}
\begin{figure}
	\vspace{-0.5em}
    \centering
    \includegraphics[width=0.40\textwidth, trim=1.0cm 1.0cm 1.0cm 1.0cm]{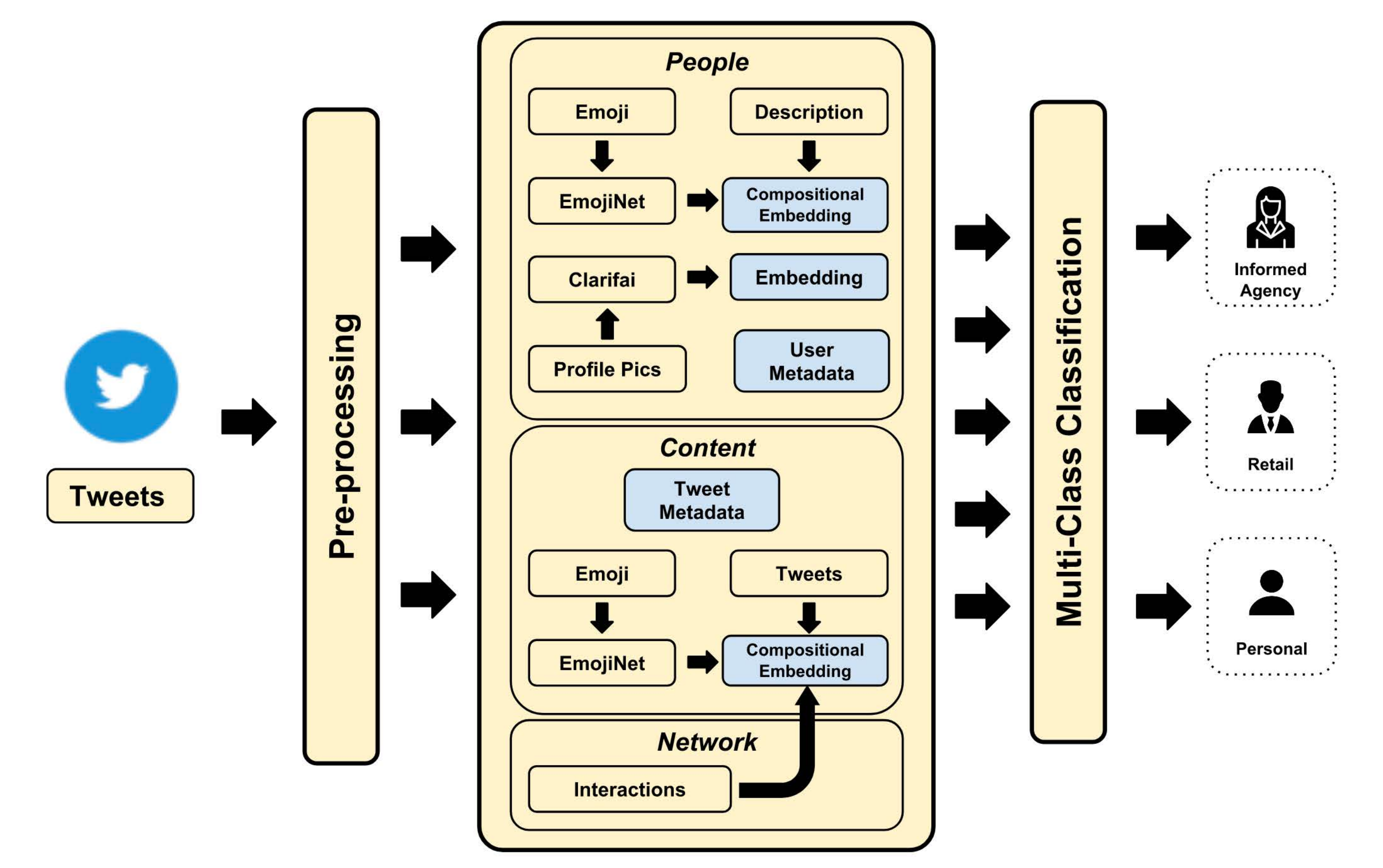}
    \vspace{-0.5em}    \caption{\footnotesize Overall Architecture and Processing showing composition of embeddings across People-Content-Network views for User Classification }
    \label{fig:arch}
 \vspace{-2em}
\end{figure}

\vspace{-0.6em}
The novelty in our approach to the user classification problem is to leverage
the multiview aspect of the Twitter data by creating compositions of embeddings for different views. As depicted in the overall architecture of our approach in Fig \ref{fig:arch}, this section provides details of critical steps in our approach. 

\vspace{-0.6em}
\subsection{Preprocessing}
At this stage, we trained two Word Embedding(WE) models for \emph{Content} and \emph{People} views using our domain-specific Twitter corpus. (i) The Content WE model is based on ~1.8M unique pre-processed tweets, and (ii) The People WE model is based on pre-processed user descriptions of ~1M unique users. We built such two separate WE models because we observed that user descriptions were more complete and contained less jargon and slang terms as compared to tweets. 

To obtain discriminative features for user classification, we removed stop words, punctuations, and alphanumeric characters from tweets and user descriptions. We also extracted URLs, mentions of screen names, retweeted user screen names, contact information (e.g., phone number, email and web address), and emoji. After that, we lemmatized the tweets and user descriptions corpus. Moreover, we employ EmojiNet \cite{wijeratne2017emojinet} to retrieve senses and keywords from emoji, and Clarifai\footnote{\url{https://www.clarifai.com/demo}}  to process profile pictures.  The overall goal is to enable gleaning of semantically relevant information about users from their tweets for reliable determination of user types.
\vspace{-0.6em}
\subsection{Correlation Analysis}
\label{sec:CA}
In this study, we perform correlation analysis between embeddings of features from different views to assess which compositional operation is appropriate. The similarity between embedding vectors derived from the textual representation of features constrains the operations that can be used to combine them since the resulting vector needs to be representative of the components. For example, when two embedding vectors are highly uncorrelated, dimensionality reduction does not generate representative vector space. However, uncorrelated embeddings can be composed merely with vector addition, to make resulting vector space more representative. 

For instance, researchers \cite{goikoetxea2016single} made use of operations such as addition and concatenation, to combine word embedding vectors of the input text. These word embeddings were generated from text corpora and knowledge bases for more contextually rich representation of the input text. 
Similarly, \cite{faruqui2014retrofitting} retrofits word vectors, using the WordNet embeddings to enrich the word embeddings of the input text. 

The creation of embedding vectors is performed through probabilistic calculations \cite{bamler2017dynamic}, and the embedding of each view (Section \ref{sec:CME}) may or may not correlate with that of the other views.  



\begin{table}[!htbp]
\vspace{-1.5em}
\centering
\footnotesize
\caption{\scriptsize Spearman ($\rho$) Correlation Analysis for View Pairs}
\vspace{-0.5em}
\begin{tabular}{|p{3.5cm}|c|c|}
\hline
\textbf{View Pairs} & \textbf{$\rho$} & \textbf{p-value} \\
\hline
User Description \& Emoji & 0.002 & $<$ 0.01  \\ \hline
Tweets \& Emoji & 0.02 & $<$ 0.01  \\ \hline
Tweets \& Network & 0.04 & $<$ 0.01  \\ \hline
User Description \& Network & 0.0001 & $>$ 0.01  \\
\hline
\end{tabular}
\label{tab:kruskal-spearman}
\vspace{-1.6em}
\end{table}

We conducted correlation analysis between different pairs of view embedding vectors as shown in Table \ref{tab:kruskal-spearman}. 
The table shows Spearman correlation and their corresponding p values for these pairs. 


We use Spearman as our correlation metric to measure the similarity between view embeddings at each data point since our embeddings do not follow the Gaussian distribution. In this analysis, our alternative hypothesis ($H_1$) is that the two embedding vectors are uncorrelated, and similarly the null hypothesis ($H_0$) is that they are correlated. Having the p-value, less than 0.01 suggests the rejection of $H_0$. Hence, based on Spearman, we see from the Table \ref{tab:kruskal-spearman} that for the first three pairs the null hypothesis of correlation $H_0$ can be rejected, while for the pair  User description and Network, we are unable to reject the null hypothesis of correlation ($H_0$).  In fact, the data indicates that people interact closely based on their similar user characteristics rather than the shared tweet content in marijuana-related communications.


\vspace{-0.5em}
\subsection{Feature Engineering}
\label{sec:FE}
\vspace{-0.2em}
In our analysis, we have organized our features under three main categories: Person, Content, and Network, since we consider these as the main views of the Twitter communication that contribute to the context.

\begin{figure}[!htbp]
  \begin{center}
  \includegraphics[width=0.4\textwidth,trim=0.0cm 0.0cm 4.0cm 2.0cm]{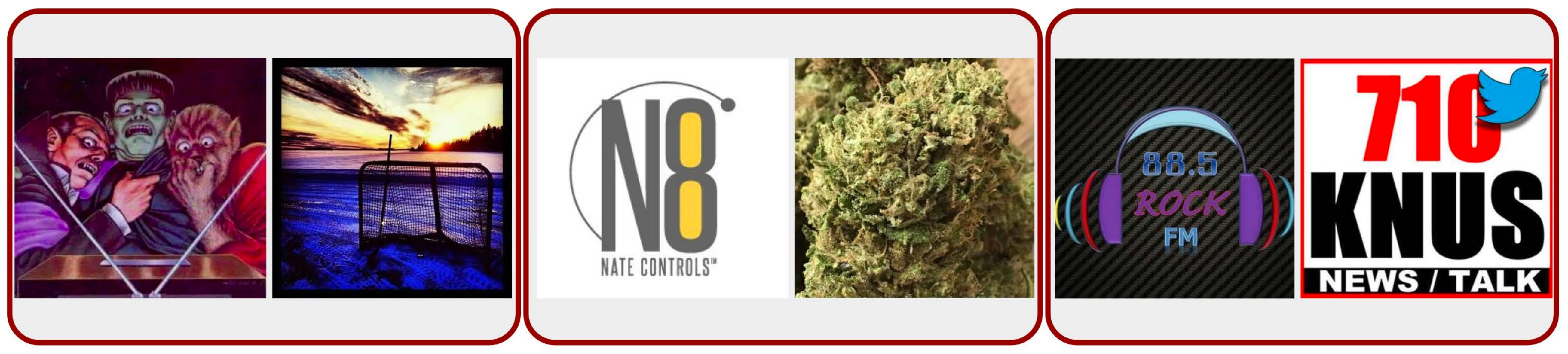}
  \end{center}
  \vspace{-1.6em}
  \caption{Example profile pictures(2 each) of P(left), R(center) and I(right)}
  \label{fig:pp}
  \vspace{-1em}
\end{figure}

\subsubsection{People}
This set of features are user-level that contributes to differentiating the user types from each other on social media. Specifically, it includes user descriptions, name, screen name, contact information and profile pictures.
\begin{itemize}
\item User Descriptions: This field holds the description of the account that was defined by the user. As this metadata carries information on characteristics of the user, we exploit the elements of this feature such as text, emoji, and contact information by employing text processing techniques. 
\item Name: This field holds the name of each user where users can enter their full personal, business, or organization name, or have an arbitrary entry. We use this information to discriminate person users utilizing a lexicon\footnote{\url{https://goo.gl/8MY5Cz}} of commonly used person's first and last names.
In fact, we found that 68\% of the person users can be identified using names listed in the lexicon.
\item Contact Information: We extract this information from the description of users as it includes a phone number, email and web addresses. Usually, retail accounts provide this information in their profile for their customers to reach out to them, making this feature a discriminative factor in classification. 
\item Profile Pictures: This visual form of Twitter data can reflect feelings, emotions, intentions, and other characteristics of a user. We consider this feature as discriminative as there is a noticeable difference in profile pictures of personal, retail, and informed agency accounts. See Fig \ref{fig:pp} for examples.
\end{itemize}  

\subsubsection{Content}
To glean  discriminatory features from tweet content, we first separated text, emoji and URLs, and then processed them separately.

\begin{itemize}
\item Tweet text: We first extracted tweet text, by filtering other elements such as mentions, URLs, and emoji, and concatenate tweets of each user. Then we created word embedding vectors out of this textual data.  
\item URLs: Users usually provide URLs in their character-limited tweets to refer to a more detailed version of their stories. For instance, retail and media accounts use URLs in their tweets to direct clients to their web page, more often than personal accounts. The number and frequency of URLs in a tweet can help to discriminate among user types.   
\item Emoji: The use of emoji provides a concise and precise expression of opinions, reactions, sentiments, and emotions concerning a topic of discussion. It is a discriminative feature in our study capturing the number and senses of emoji used by different user types. 
\end{itemize}

\subsubsection{Network} 
As users on Twitter primarily interact using replies, mentions, and retweets, we utilize these interactions as our features to identify communication patterns for each user type. We consider replies as mentions.  In our exploratory data analysis of marijuana-related communications, we found that the following features are prominent. 

\begin{itemize}
\item Mentions: It is a derived feature where the author mentions the screen-name of another user and is considered as direct interaction. 
\item Retweets: It is a derived feature where the retweeting user forwards another user’s tweet and is considered a direct interaction between these two users.
\end{itemize}
We generate network embeddings by creating the adjacency matrix based on these interactions between users. This procedure is further explained in Section \ref{sec:NE}

\vspace{-0.4em}
\subsection{Compositional Multiview Embedding (CME)}
\label{sec:CME}

The Twitter data contains multiple dimensions that we call views, such as People, Content, and Network.
These views can be leveraged to contextualize a comprehensive and multi-level analysis of the Twitter social network. 

In our study, we employed the Content and People WE models for generating embeddings for Content view (e.g., Tweets) and People view (e.g., User Descriptions and Profile Pictures), respectively. 

  
As described in Section \ref{sec:enet}, the tweet content and user descriptions involve emoji, which we regard as critical for interpreting the meaning. For this reason, we extracted the textual representation of emoji from EmojiNet, and generated cumulative emoji embeddings utilizing a pre-trained word embedding model that was trained over Wikinews corpus\cite{mikolov2017advances} as explained in \cite{wijeratne2017semantics}. We also generated word embeddings for profile pictures of users. As Clarifai provides a set of tags that textually represents the profile images, we input these tags into the People WE model because we consider profile pictures as related to the People view. Then we generated CMEs by combining the embeddings at the intersection of different views of the Twitter data, as formulated below.

For Person and Content views (T), word embedding vector ($WV$) in each data point ($WV_{T_i}$, \emph{i} represents an index of a data point in a view) is calculated by averaging the word-vectors of each word that is present in the view. For instance, we preprocess the tweets of a user and generate word vectors of each word in 300 dimensions. Then we sum these vectors and divide by the number of words to generate the embedding vector for tweets of the user. However, while we perform the average operation to generate embedding vectors for Person and Content views, we do not perform average for the Network view. For generation of network embeddings, we utilized interactional features (mentions and retweets) and performed t-SVD to generate dense embeddings, where each embedding has 300 dimensions. The procedure is detailed in Section \ref{sec:NE}. 


We formally define the calculation of $WV_{T_i}$ as
$\mathbf{WV_{T_i}} = \frac{\sum_{w \in T_i \cap V} \vec{\mathbf{v}}_w}{|T_i \cap V|}$,
where $\vec{\mathbf{v}}_w$ is the embedding of word \textbf{w} and {\it V\/} 
is the vocabulary of the Content WE model trained over the marijuana-related tweet corpus.  

\begin{figure}[htp]
  \centering
  {\includegraphics[width=0.4\textwidth, trim=3.0cm 1.0cm 3.0cm 1.7cm]{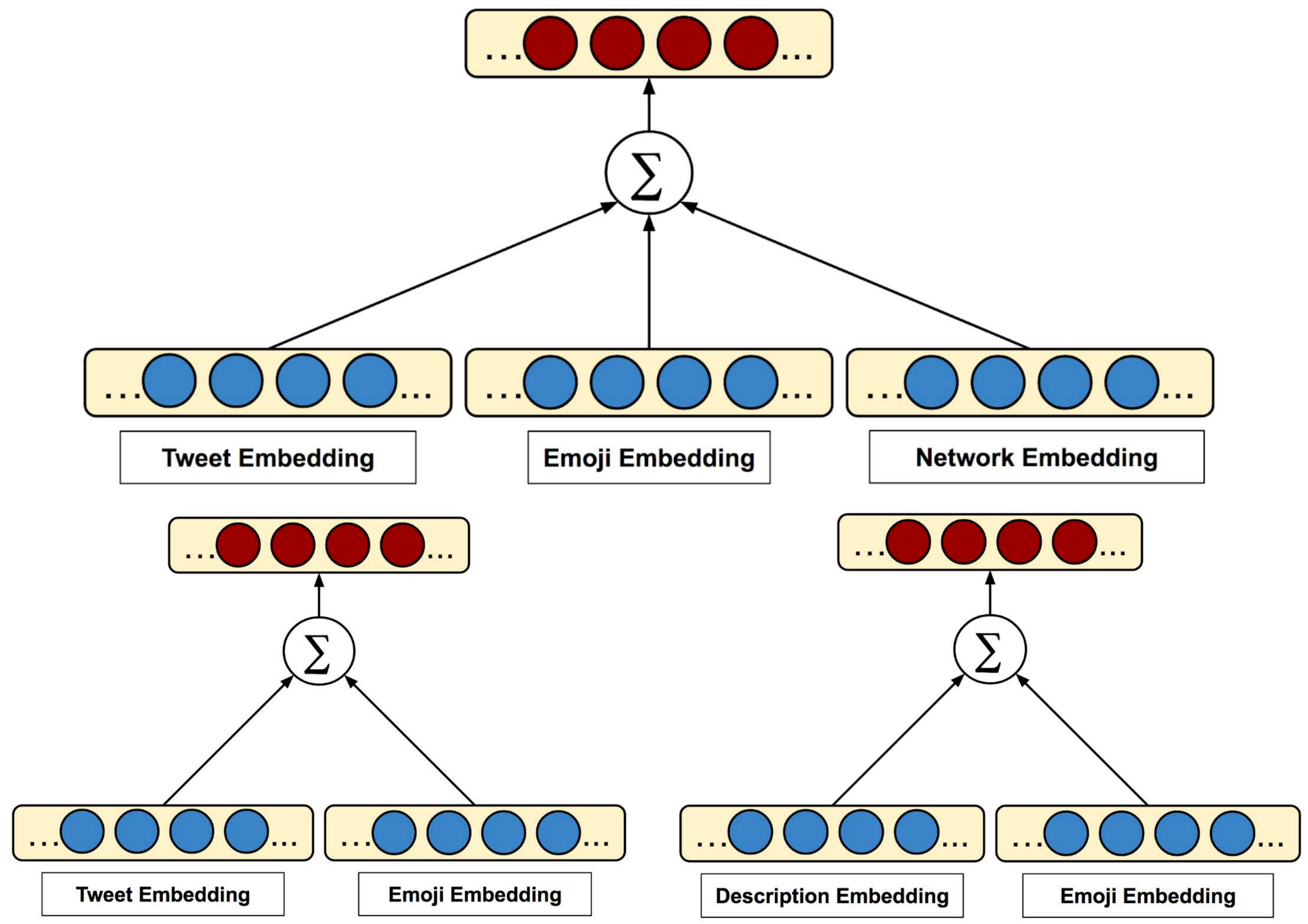}}\quad
  \vspace{-0.5em}
  \caption{Creation of CMEs for Tweet, Description, Emoji and Network}
 \label{fig:eda}
\vspace{-0.75em}
\end{figure}

\subsubsection{Tweet-Emoji(T+E) \& User Description-Emoji(D+E)}
We explained the procedure for generating WEs for Tweets, User Descriptions and Emoji earlier in this section, and we explain here how we generate CMEs for Tweet \& Emoji, and User Description \& Emoji.

As depicted in Fig. \ref{fig:eda}, to generate the Tweet-Emoji CME, we combine the WE vectors, which we generated for Tweets and Emoji, by performing the \emph{vector addition} operation. Similarly for User Description-Emoji CME, we combine the WE vectors for User Descriptions and Emoji via the vector additon. 

\subsubsection{Network Embedding (N)}
\label{sec:NE}
The user types that we characterize in this study have different volumes of network activities. For instance, while average retweet and mention rates (derived from Table \ref{tab:desc}) per user are 0.9 and 0.09 respectively on personal accounts, they are  11.08 and 3.53 on informed agency accounts. Clearly, the network activity can be used to distinguish and recognize these user types.
Thus,  combining  the network activity information with tweet content and user information can  contribute to a reliable classification. 

For representing the network activities of users, we created the weighted adjacency matrix of interactions; however, the adjacency matrix was sparse that made the generalization of the classifier difficult. 
Hence, creating dense vectors is imperative for better representation. For generating a low dimensional dense vector, we utilize truncated Singular Value Decomposition (t-SVD) which has proven to generate dense embedding in NLP and network embedding tasks \cite{tsitsulin2018verse}. 

Formally, we define the adjacency matrix as $\mathbf{A}\in \mathbb{R}^{m\times n}$, where $\mathbf{m}$ and $\mathbf{n}$ denote source users and target users respectively (capturing  direction of communication). 
\begin{equation}
    A_{u_i,u_j} = Interaction Count_{u_i,u_j}
\vspace{-0.5em}
\end{equation}
where, for a pair of users $u_i, u_j$, $A_{u_i,u_j}$ represents a cell in the matrix $\mathbf{A}$ of dimension $|m|\times|n|$ representing interaction counts, which includes both retweets and mentions, for the corresponding users. 

The adjacency matrix $\mathbf{A}$ is sparse and non-stochastic ($\sum_{j=1}^n A_{i,j} \neq 1$). As we need to create a dense and stochastic representation of the network activities, we normalize the values in a row such that they will all sum up to 1. This normalization is done by diving every value by the sum of all values in a row, and this process makes the matrix stochastic. As the number of sources and target users is mostly different in $\mathbf{A}$ ($1149\times1701$), we convert $\mathbf{A}$ to square cosine matrix, denoted by $\mathbf{A^{cosine}} \in \mathbb{R}^{m \times m}$ obtaining a matrix $1149\times1149$, since we want to measure the similarity between users in our training set. Transformation of $\mathbf{A}$ to $\mathbf{A^{cosine}}$ is formulated as follows:
$A^{cosine} = \frac{A \cdot A^T}{||A||||A^T||}$. Each cell value in $\mathbf{A^{cosine}}$ lies between 0 and 1 and is symmetric. 

As our adjacency matrix $\mathbf{A^{cosine}}$ is  $1149\times1149$, we need to reduce its dimension down to 300 for us to perform composition of the network embedding with other word embeddings. Therefore, we apply t-SVD over the matrix $\mathbf{A^{cosine}}$ resulting three square matrices: U, $\textSigma$, $U^T$ $\in$ $\mathbb{R}^{m \times m}$, where $\textSigma = \{ \sigma_1, \sigma_2, ... , \sigma_m\}$ is a set of $\mathbf{m}$ singular values. After we apply the dimensionality reduction, the reduced matrix becomes of dimension $m \times 300$. We denote the reduced matrix as $\mathbf{A^{reduced}} \in \mathbb{R}^{m \times 300}$ and  its value is determined by: $ \mathbf{A^{reduced}} =  U_{m \times 300} \cdot (\textSigma_{300 \times 300}^{-1})^T$. 

The 300 dimensional embeddings in $ \mathbf{A^{reduced}}$ is considered as the network embedding of users, and is used to create a CME in our user type classification. 

\subsubsection{Network-Tweet-Emoji(N+T+E)}
After we generate the network embeddings(NE) of users, we combine the WEs for Tweets and User Descriptions, and NE to generate the Network-Tweet-Emoji CME by performing the \emph{vector addition} operation. Embeddings for Network, Tweets and Emoji are all in 300 dimensions.
\vspace{-0.65em}
\subsection{Experimental Setting} 
In building the Content and Person WE models, we used Skip-gram model with negative sampling. The rate of negative sampling was set to 10 and the window size was set to 5. Such a set up is desirable for datasets of average-size \cite{mikolov2013distributed}. The Content WE model was trained on a pre-processed corpus of ~1.8M unique tweets generated from ~1M unique users creating a vocabulary (V) of 16,531 words. The People WE model was trained on 946,975 pre-processed user descriptions obtained from ~1M unique users, generating a vocabulary (V) of 16,903 words. Apart from linguistic differences between user descriptions and tweets, another reason to build two WE models is multiview aspect of our dataset that also includes profile pictures and emoji in a profile that reflect different contextual meanings as compared to the tweets of a user. In order to create an embedding of a profile picture, we used Clarifai to generate text caption and then apply the Person WE model on the text caption. \\
\textbf{Empirical Baseline:} To the best of our knowledge, the problem of user type classification in marijuana-related communications on Twitter that we address in this study has not been investigated before. For this reason, we created an empirical baseline that utilizes word embeddings of the textual content of tweets and descriptions.\\ 
We conducted two sets of experiments depending on the inclusion of CME with network level features. The first set of experiments do not include the CME with network level features, and we incrementally include the Person and Content level features. We utilize all data points in our training set that comprises of 4982 users. As discussed in earlier sections, interactions also play an essential role in forming the characteristics of user types. Therefore, the second set of experiments included CMEs which contain Network level features, where we take the best performing classification setting from the first set of experiments as a baseline for comparison. At this stage, we had to reduce the size of the training set down to 1149 users where the sizes of P, I, and R classes were 1045, 87 and 17 respectively. 
Since our training set was highly imbalanced, we applied the oversampling algorithm SMOTE to avoid bias towards the majority class at the expense of  the minority classes. 

In our experiments, to illustrate the improvement that the domain specific WE models provides, we also utilized a generic word2Vec model, called Tweet2Vec \cite{godin2015multimedia}, for a comparison, which is explained in detail in Section \ref{sec:R}. 
\\

\vspace{-0.5em}
\section{RESULTS}
\label{sec:R}
\vspace{-0.5em}
Table\ref{tab:result1} and Table \ref{tab:results2} present the results of the two sets of experiments. The first set of experiments involve only user profile and tweet content level features, whereas the second set of experiments involve the addition of network features. To illustrate the improvement obtained by the addition of network level features into the classification, we take the best performing approach of the first set of experiments as the baseline for the second set of experiments.
 
The different feature sets incorporate different views of the data as explained in Section \ref{sec:M}. We systematically and gradually include person-content-network features to observe their individual contributions to the outcome of the classification.  

\begin{table}[!htbp]
\vspace{-1em}
\centering
\caption{\footnotesize Results on Classification of User Types with 4982 Users.}
\vspace{-0.5em}
\resizebox{0.5\textwidth}{!}{
\begin{tabular}{|p{1.8cm}|c|c|c|c|c|c|c|c|c|p{0.5cm}|}
\hline
\multirow{2}{*}{\textbf{Feature Set}} & \multicolumn{3}{c|}{\textbf{Precision}} & %
    \multicolumn{3}{c|}{\textbf{Recall}} & \multicolumn{3}{c|}{\textbf{F-score}} & \multirow{2}{*}{\textbf{Avg.F}}\\
\cline{2-10}
 & \textbf{P} & \textbf{I} & \textbf{R} & \textbf{P} & \textbf{I} & \textbf{R} & \textbf{P} & \textbf{I} & \textbf{R} & \\
\cline{1-11}
E(T),E(D) & 0.91 & 0.86 & 0.79 & 0.99 & 0.27 & 0.67 & 0.95 & 0.42 & 0.73 & 0.88 \\ \cline{1-11}
T2V(T),T2V(D) & 0.89 & 0.87 & 0.87 & 0.99 & 0.10 & 0.66 & 0.94 & 0.18 & 0.75 & 0.86 \\ \cline{1-11}

E(T+E),E(D+E) & 0.89 & 0.96 & 0.88 & 0.99 & 0.10 & 0.60 & 0.94 & 0.18 & 0.71 & 0.85 \\ \cline{1-11} 

E(T+E),E(D+E), &&&&&&&&&&\\
TMD,UMD & 0.89 & 0.95 & 0.84 & 0.99 & 0.09 & 0.63 & 0.94 & 0.17 & 0.72 & 0.85 \\ \cline{1-11}

E(T+E),E(D+E), &&&&&&&&&&\\
TMD,UMD,PP & 0.97 & 0.99 & 0.88 & 0.99 & 0.77 & 0.92 & 0.98 & 0.87 & 0.90 & 0.97 \\ \cline{1-11}

\hline
\end{tabular}}
\label{tab:result1}
\end{table}

\begin{table}[!htbp]
\vspace{-2em}
\centering
\caption{\footnotesize Results for Classification of User Types with 1149 Users}
\vspace{-0.5em}
\resizebox{0.5\textwidth}{!}{\begin{tabular}{|p{1.7cm}|c|c|c|c|c|c|c|c|c|c|}
\hline
\multirow{2}{*}{\textbf{Feature Set}} & \multicolumn{3}{c|}{\textbf{Precision}} & %
    \multicolumn{3}{c|}{\textbf{Recall}} & \multicolumn{3}{c|}{\textbf{F-score}} & \multirow{2}{*}{\textbf{Avg.F}}\\
\cline{2-10}
 & \textbf{P} & \textbf{I} & \textbf{R} & \textbf{P} & \textbf{I} & \textbf{R} & \textbf{P} & \textbf{I} & \textbf{R} & \\
\cline{1-11}

E(T+E),E(D+E), &&&&&&&&&&\\
TMD,UMD,PP & 0.96 & 0.98 & 0.93 & 0.99 & 0.57 & 0.82 & 0.97 & 0.72 & 0.87 & 0.95 \\ \cline{1-11}

E(N),E(T+E), &&&&&&&&&&\\
E(D+E),TMD &&&&&&&&&&\\
UMD,PP & 0.95 & 0.95 & 0.95 & 1.0 & 0.52 & 0.80 & 0.97 & 0.67 & 0.87 & 0.95 \\ \cline{1-11}

E(N+T+E), &&&&&&&&&&\\
E(D+E),TMD &&&&&&&&&&\\
UMD,PP & 0.96 & 0.98 & 0.97 & 1.0 & 0.58 & 0.86 & 0.98 & 0.73 & 0.91 & 0.96 \\
\hline
\end{tabular}}
\textbf{E:Embedding, T:Tweet, D:Description, N:Network, T2V:Tweet2Vec, TMD:Tweet Metadata, UMD:User Metadata, PP: Profile Pictures}
\label{tab:results2}
\vspace{-1.2em}
\end{table}

We evaluated our approach using Average F-score (Avg.F) for each user type (P,I,R).
We also report precision, recall, and average F-score, and discuss the overall performance. 

The baseline approach that we empirically chose achieved an overall F-score of 88\% using the word embeddings of tweets content and user descriptions. The F-scores for individual classes of P, I, and R were 95\%, 42\%, and 73\%, respectively. We generated these embedding vectors using the domain-specific word embedding models.\\ 
As we see in Table \ref{tab:result1} that the classifier built with the embeddings of tweets and descriptions generated through the Tweet2Vec model obtained an average  F-score of 86\%, and underperformed for P and  I classes. Therefore, we continued experiments using Content and People WE models. 

As discussed in Section \ref{sec:M}, to better contextualize different elements of the content such as text and emoji, we have generated CMEs from the tweets and emoji embeddings, and similarly from user descriptions and emoji. Though this experiment has shown a reduction of average F-score by 3\%, the precision has been improved by 10\% for I and R classes, meaning false positives for I and R are reduced. Given the small size of these classes in our training dataset, such improvement in precision encouraged us to further continue our experiments with the inclusion of CMEs.

We have further included the tweet and user metadata to the feature set, and it still did not make a significant difference in the performance. However, the inclusion of profile pictures as a feature in the experiments showed a significant improvement in the overall F-score to 97\%, where F-scores for P, I, and R were 98\%, 87\%, and 90\%, respectively. As discussed earlier, we can benefit from the multiview aspect of the Twitter data to cultivate more satisfactory interpretation of the content. The inclusion of textual data, emoji and profile pictures in our approach by combining them through CMEs for classification of user types, has impacted the outcome significantly.
Furthermore, recall that, in the second set of experiments, we have extended our study by applying our approach with the addition of network interactions between users. We have generated network embeddings from the interactions between users. We have used the best performing classifier from the first set of experiments(Table \ref{tab:result1}) as a baseline for the second set of experiments, to compare our approach that incorporates the network embeddings.

In our second set of experiments, we have first added the network embedding as a separate feature along with the features from the second baseline approach, and it did not affect the performance. Then we created CME from the embeddings of tweets, emoji, and network, and it boosted the performance of each class, P, I, and R in terms of their F-scores, by 1\%, 6\%, and 4\%, respectively. It also improved the overall F-score by 1\%. The improvement that we achieved by applying CMEs is significant since the F-score for the second baseline was already significantly high, and our approach has improved upon that performance.

\section{CONCLUSION and FUTURE WORK}
\label{sec:C}
\vspace{-0.5em}
Our overarching goal was to utilize people, content, and network related features in marijuana-related communications on Twitter to classify the user types into three prominent categories: Personal, Informed Agency, and Retail accounts. Such a classification provides support for understanding the dynamics of issues related to marijuana and its variants from location and temporal perspectives ultimately. Furthermore, dominant and trending topics can be identified for each user type for more precise and reliable subjective analysis of related events and their impacts.

In this paper, we introduced an approach to classify user types utilizing Compositional Multiview Embedding (CME). For this purpose, we learned a domain-specific embedding for tweet text, a separate embedding for user profile descriptions, and a mapping of profile images to tags to obtain their embeddings, while incorporating emojis as words using EmojiNet embeddings. We also incorporated interactional features by creating network embeddings. Overall, we achieved 7\% improvement over the empirical baseline, when we used the CMEs without network embedding and 8\% improvement when we used the CMEs with network embedding. The latter also resulted in an F-score of 0.96. 


Although we are implicitly addressing the homophily through assessing the similarity between users based on different views, we plan to enhance our work by analyzing homophily in marijuana-related communications on Twitter as a case study by leveraging the approach explained in this paper. Upon the completion of review process, we will outsource our baseline and annotated dataset for reproducibility.

\section*{ACKNOWLEDGEMENT}
\footnotesize
Research reported in this publication was supported by National Institute on Drug Abuse (NIDA) of the National Institutes of Health (NIH) under award number 5R01DA039454-03. The content is solely the responsibility of the authors and does not necessarily represent the official views of the NIH.


\footnotesize
\bibliography{reference}

\begin{thebibliography}{10}
\providecommand{\url}[1]{#1}
\csname url@samestyle\endcsname
\providecommand{\newblock}{\relax}
\providecommand{\bibinfo}[2]{#2}
\providecommand{\BIBentrySTDinterwordspacing}{\spaceskip=0pt\relax}
\providecommand{\BIBentryALTinterwordstretchfactor}{4}
\providecommand{\BIBentryALTinterwordspacing}{\spaceskip=\fontdimen2\font plus
\BIBentryALTinterwordstretchfactor\fontdimen3\font minus
  \fontdimen4\font\relax}
\providecommand{\BIBforeignlanguage}[2]{{%
\expandafter\ifx\csname l@#1\endcsname\relax
\typeout{** WARNING: IEEEtran.bst: No hyphenation pattern has been}%
\typeout{** loaded for the language `#1'. Using the pattern for}%
\typeout{** the default language instead.}%
\else
\language=\csname l@#1\endcsname
\fi
#2}}
\providecommand{\BIBdecl}{\relax}
\BIBdecl

\bibitem{kursuncu}
U.~Kursuncu, M.~Gaur, U.~Lokala, K.~Thirunarayan, A.~Sheth, and I.~B. Arpinar,
  ``{Predictive Analysis on Twitter: Techniques and Applications},'' in
  \emph{Springer-Nature}, 2018.

\bibitem{benton2016learning}
A.~Benton, R.~Arora, and M.~Dredze, ``Learning multiview embeddings of twitter
  users,'' in \emph{ACL}, 2016.

\bibitem{mitchell2010composition}
J.~Mitchell and M.~Lapata, ``Composition in distributional models of
  semantics,'' \emph{Cognitive science}, 2010.

\bibitem{zhang2016user}
D.~Zhang, S.~Li, H.~Wang, and G.~Zhou, ``User classification with multiple
  textual perspectives,'' in \emph{COLING}, 2016.

\bibitem{rizoslearning}
G.~Rizos, S.~Papadopoulos, and Y.~Kompatsiaris, ``Learning to classify users in
  online interaction networks.''

\bibitem{liao2017attributed}
L.~Liao, X.~He, H.~Zhang, and T.~Chua, ``Attributed social network embedding,''
  \emph{arXiv preprint arXiv:1705.04969}, 2017.

\bibitem{pennacchiotti2011democrats}
P.~A. Pennacchiotti, M, ``Democrats, republicans and starbucks afficionados:
  user classification in twitter,'' in \emph{ACM SIGKDD}, 2011.

\bibitem{bergsma2013broadly}
S.~Bergsma, M.~Dredze, B.~Van~Durme, T.~Wilson, and D.~Yarowsky, ``Broadly
  improving user classification via communication-based name and location
  clustering on twitter,'' in \emph{NAACL-HLT}, 2013.

\bibitem{campbell2013content+}
W.~Campbell, E.~Baseman, and K.~Greenfield, ``Content+ context networks for
  user classification in twitter,'' in \emph{NIPS}, 2013.

\bibitem{de2012unfolding}
M.~De~Choudhury, N.~Diakopoulos, and M.~Naaman, ``Unfolding the event landscape
  on twitter: classification and exploration of user categories,'' in \emph{ACM
  CSCW}, 2012.

\bibitem{tinati2012identifying}
R.~Tinati, L.~Carr, W.~Hall, and J.~Bentwood, ``Identifying communicator roles
  in twitter,'' in \emph{WWW}, 2012.

\bibitem{colleoni2014echo}
E.~Colleoni, A.~Rozza, and A.~Arvidsson, ``Echo chamber or public sphere?
  predicting political orientation and measuring political homophily in twitter
  using big data,'' \emph{Journal of Communication}, 2014.

\bibitem{fang2015topic}
A.~Fang, I.~Ounis, P.~Habel, C.~Macdonald, and N.~Limsopatham, ``Topic-centric
  classification of twitter user's political orientation,'' in \emph{ACM
  SIGIR}, 2015.

\bibitem{kim2017classification}
A.~Kim, T.~Miano, R.~Chew, M.~Eggers, and J.~Nonnemaker, ``Classification of
  twitter users who tweet about e-cigarettes,'' \emph{JMIR}, 2017.

\bibitem{purohit2011understanding}
H.~Purohit, Y.~Ruan, A.~Joshi, S.~Parthasarathy, and A.~Sheth, ``Understanding
  user-community engagement by multi-faceted features: A case study on
  twitter,'' in \emph{WWW Workshop SoME}, 2011.

\bibitem{purohit2015intent}
H.~Purohit, G.~Dong, V.~Shalin, T.~Prasad, and A.~Sheth, ``Intent
  classification of short-text on social media,'' in \emph{Smart
  City/SocialCom/SustainCom (SmartCity), 2015 IEEE International Conference
  on}.\hskip 1em plus 0.5em minus 0.4em\relax IEEE, 2015.

\bibitem{wijeratne2017emojinet}
S.~Wijeratne, L.~Balasuriya, A.~Sheth, and D.~Doran, ``Emojinet: An open
  service and api for emoji sense discovery,'' \emph{arXiv preprint
  arXiv:1707.04652}, 2017.

\bibitem{mikolov2013distributed}
T.~Mikolov, I.~Sutskever, K.~Chen, G.~Corrado, and J.~Dean, ``Distributed
  representations of words and phrases and their compositionality,'' in
  \emph{NIPS}, 2013.

\bibitem{lilleberg2015support}
J.~Lilleberg, Y.~Zhu, and Y.~Zhang, ``Support vector machines and word2vec for
  text classification with semantic features,'' in \emph{IEEE ICCI* CC}, 2015.

\bibitem{wang2016semantic}
P.~Wang, B.~Xu, J.~Xu, G.~Tian, C.~Liu, and H.~Hao, ``Semantic expansion using
  word embedding clustering \& convolutional neural network for improving short
  text classification,'' \emph{Neurocomputing}, 2016.

\bibitem{godin2015multimedia}
F.~Godin, B.~Vandersmissen, W.~De~Neve, and R.~Van~de Walle, ``Multimedia lab
  $@ $ acl wnut ner shared task: Named entity recognition for twitter
  microposts using distributed word representations,'' in \emph{Proceedings of
  the Workshop on Noisy User-generated Text}, 2015.

\bibitem{sheth2016semantic}
A.~Sheth and P.~Kapanipathi, ``Semantic filtering for social data,'' \emph{IEEE
  Internet Computing}, 2016.

\bibitem{goikoetxea2016single}
J.~Goikoetxea, E.~Agirre, and A.~Soroa, ``Single or multiple? combining word
  representations independently learned from text and wordnet.'' in
  \emph{AAAI}, 2016.

\bibitem{faruqui2014retrofitting}
M.~Faruqui, J.~Dodge, S.~K. Jauhar, C.~Dyer, E.~Hovy, and N.~Smith,
  ``Retrofitting word vectors to semantic lexicons,'' \emph{arXiv preprint
  arXiv:1411.4166}.

\bibitem{bamler2017dynamic}
R.~Bamler and S.~Mandt, ``Dynamic word embeddings,'' in \emph{ICML}, 2017.

\bibitem{mikolov2017advances}
T.~Mikolov, E.~Grave, P.~Bojanowski, C.~Puhrsch, and A.~Joulin, ``Advances in
  pre-training distributed word representations,'' \emph{arXiv preprint
  arXiv:1712.09405}, 2017.

\bibitem{wijeratne2017semantics}
S.~Wijeratne, L.~Balasuriya, A.~Sheth, and D.~Doran, ``A semantics-based
  measure of emoji similarity,'' \emph{arXiv preprint arXiv:1707.04653}, 2017.

\bibitem{tsitsulin2018verse}
A.~Tsitsulin, D.~Mottin, P.~Karras, and E.~M{\"u}ller, ``Verse: Versatile graph
  embeddings from similarity measures,'' in \emph{WWW}, 2018.

\end{thebibliography}
\bibliographystyle{IEEEtran}

\end{document}